\documentstyle[epsf]{l-aa}
\begin{document}
\thesaurus{13.07.1} 
\title{On the possible origin of gamma ray bursts as a result of
       interaction of relativistic jets with the soft photon field
       in dense stellar regions}

\author{ Darja N. Drozdova\inst{1}     \and
       Ivan E. Panchenko  \inst{1}
       }
\date{}
\institute{
 Department of Physics, Moscow State University
}
\maketitle

\markboth{D. Drozdova, I. Panchenko: GRB in dense stellar regions}{}
\begin{abstract}
 We examine and develop the model of  gamma ray bursts
 origin proposed by Shaviv and Dar (1996), according to which
 the strong gamma ray emission is produced by the interaction of
 the baryonic relativistic jet with the soft photon field
 in a dense stellar region.
 By the simulations of the burst profiles we show that these profiles
 are sensitive to the jet geometry.
 Also we derive and discuss the model restrictions based on the event rate.
\keywords{Gamma ray bursts -- jets -- neutron stars}
\end{abstract}

\section{Introduction}

The idea of cosmological origin of gamma ray bursts
(Prilutski \& Usov 1975; Usov \& Chibisov 1975)  is well supported by
their observed isotropy on the sky and non-uniform spatial
distribution  Meegan et al. 1992).
The enormous luminosity needed for explanation of gamma ray bursts
if their origin is cosmological imply that the best candidates for their
sources are the mergers of binary neutron stars and/or neutron stars with
black holes at redshifts  ($z\simeq 1-2$),
(Blinnikov, et al. 1984; Paczy\'nski 1991, 1992).
Lipunov et al (1995) showed that
the cosmological origin of gamma ray bursts is consistent with
the observed BATSE gamma ray bursts $\log N$--$\log S$ distribution, but the
beaming of the radiation is required for the balancing of the observed and
predicted event rates.

The cosmological models for gamma ray bursts have not yet been proven;
moreover, they come across severe problems, one of them being the
problem of the efficiency of transforming the gravitational binding energy
into gamma-rays. The explanation of the observed profiles of the bursts
remain another unsolved problem.

Recently Shaviv and Dar (1995,1996a,b) proposed a model of
the gamma ray bursts
origin which was the first one to reproduce the observed burst profiles,
and also solved the problem of energy transformation into gamma rays.
In this model, a gamma ray burst is produced  by reemission of the
soft field photon ($E\sim 1$~eV) by the atoms of a relativistic jet
(or expanding shell)
(with $\Gamma \sim 1000$), which transforms these photons to those of
the energies $\sim \Gamma^2E\sim 1$~MeV.

It is suggested that such a jet or shell can be generated during a binary
neutron star merger.
In a dense stellar region such as a globular cluster core or a galaxy center,
the particles of the jet would pass close to several stars.
The light photons filling the vicinity of the stars in the comoving frame of the
jet would have energy $\sim\Gamma E\sim 1$~keV.
Shaviv and Dar show that these photons could be absorbed
by photoionization or photoexcitation of heavy atoms (like iron)
that may be present in the jet, the cross section of such absorption
being much higher than that of Compton scattering.
When reemitted, in the rest frame of the
observer the photons will have energy of the order of
$\Gamma^2 E \sim 1$~Mev and will be beamed into the $1/\Gamma$ angle.
Thus, when passing by a star, the jet would produce a burst of $\gamma$
radiation.

These $\gamma$ photons generated closer to the jet source
should come to the observer earlier
than those generated far from it because the jet expansion velocity
$v=c\sqrt{1-1/\Gamma^2}$
is smaller than the velocity of light.
Therefore,
each of the stars on the way of the jet would produce a peak in the burst
profile.
Below we will determine the shape of this peak.
 In general,  the observed burst profile would
map the distribution of the soft photon density on the way of the jet:
$u(R) \rightarrow F(t)$, where $R$ is the distance from
the source to the current point, $t$ is the total time taken the signal
to pass from the jet source to the observer: $t=R/v+(D-R)/c$,
where $D$ is the distance from the source to the observer.

The above model explains many of the observed features
of gamma ray bursts, including the profiles.

\section{The geometry of the emitting region}
Let us now discuss the formation of the burst profile with taking
into account the $\gamma$ emission generated out of the line of sight.

If the jet particles moving with the relativistic velocity emit isotropically,
then in the observer frame their emission will be concentrated into a narrow
beam of the width $\theta_0\sim 1/\Gamma$ centered with the particle motion
direction.
For $\Gamma\gg 1$ the angular distribution of the emission in the observer frame
is approximated by
\begin{equation}
A(\theta) = 1/\left(1 + \Gamma^2\cos^2\theta\right)
\end{equation}

In the most general case we can determine the burst flux profile
by computing the following integral:
\begin{equation}\label{F(t)}
   F(t) = \int
     \rho(\vec R,t - \frac{|\vec R - \vec D|}{c})u(\vec R)A(\theta)d\vec R,
\end{equation}
where $\rho(\vec R,t)$ is the density of rate of the soft to hard photon
conversion  by the jet atoms in space and time,
$u(\vec R)$ is the density of soft photons  and $A(\theta(\vec R,t))$ is the
angular distribution of the emitted light in the observer frame, the integral
being taken over the whole space. $\theta(\vec R,t)$
here is the angle between the particle velocity direction and the direction
to the observer.

If we assume that the jet is optically thick (for the photons of
the energy $\Gamma E$ in its rest frame), then the we would see the emission
from a geometrically thin region (the jet ``front''). Then we can
approximate $\rho(\vec R,t)$ as
\begin{equation}
     \rho(\vec R,t)  = \delta(\vec R - \vec R(t)),
\end{equation}
where $\vec R(t)$ defines expansion of the jet front.
Shaviv and Dar assumed that the jet has a plane front moving in the direction
of the observer. Then in equation~(\ref{F(t)}) $\theta=0$ anywhere and
\begin{equation}
\vec R(t) = \vec vt + \vec R_\perp, \qquad \vec v=const.
\end{equation}
Then the observed flux would be
\begin{equation}\label{PlaneFlux}
F(t) \sim u((\frac Dc -t)\frac{v}{1-B}),
\end{equation}
where $B=v/c$, i.e. the burst profile reflects the soft photon
density in the vicinity of the jet.

Let us now study another case, when the jet front expands spherically
and then
discuss the difference between these two geometries of the jet expansion.
In our case
\begin{equation}
\vec R(t) = \vec vt, \qquad \vec v = v\vec n.
\end{equation}

\begin{figure}
\epsfxsize=6cm
\centerline{\epsfbox{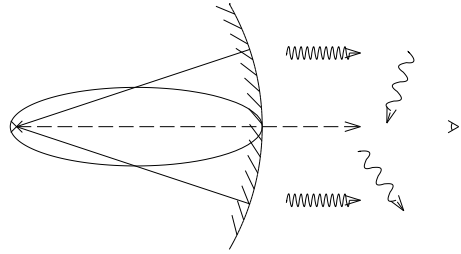}}
\caption{
A sketch of the model: We see simultaneously the $\gamma$ photons
coming from the ellipsoid expanding after the spherical jet front.
}
\label{Sketch}
\end{figure}

The main difference between the spherical and plane front is that for the
spherical one the emission from different parts of the front arrive to the
observer at different time. This effect is well known as the rings around
supernovae.
The $\gamma$ photon arrival time from the point with the
coordinates $(R,\theta)$
for the spherical front geometry is
\begin{equation}
\Delta t = \frac Rv + \frac {(D-R\cos\theta)}{c}.
\end{equation}
We detect the $\gamma$ photons from the surface with $\Delta t = const$
simultaneously. For the plane jet front, this surface coincides with the
front
while for the case of the spherical front this surface is an ellipsoid
with the jet expansion source in a focus:
\begin{equation}
 R = \frac{vt-DB}{1-B\cos\theta}.
\end{equation}
This ellipsoid is strongly elongated in the direction of the observer --
its eccentricity is as close to unity as $B\approx0.999995$.
Due to the fact that the jet $\gamma$-emission is concentrated in a narrow
beam with width $\theta_0\approx 1/\Gamma$, we will see the emission
only from the part of the ellipsoid located inside the $1/\Gamma$ cone
with the vertex at the jet source.
It is easy to show that the ellipsoid
and the cone intersect at $R_{min} \approx 0.5 R_{max}$, where
$R_{max}$ is the apocenter distance of the ellipsoid (Fig.~\ref{Sketch}).

\begin{figure*}
\epsfxsize=10cm
\centerline{\epsfbox{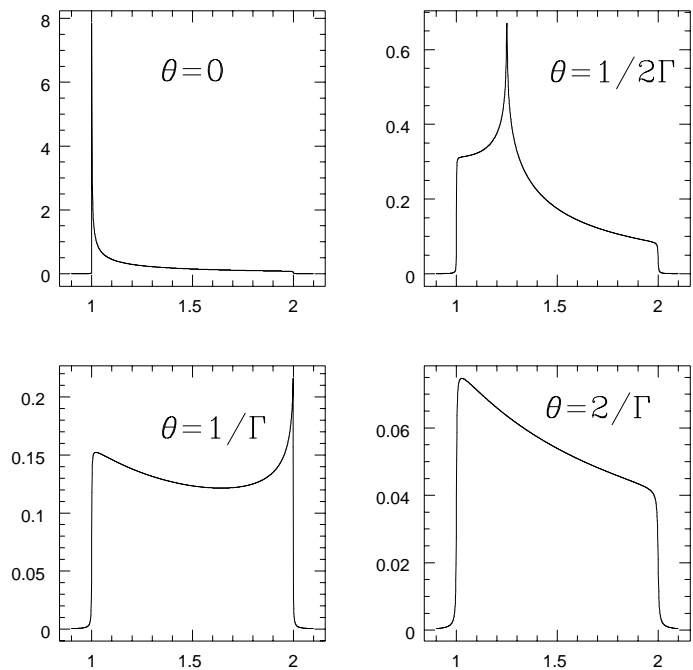}}
\caption{The sample burst profile produced by a single star at different
angular distances from the explosion-observer line:
$\theta=0,0.5/\Gamma,1/\Gamma,2/\Gamma$. The flux is shown in arbitrary units;
time is in the units of $R(1-B)/c$.
}
\label{OneStarProfiles}
\end{figure*}

Thus
we detect simultaneously the emission coming from rather big region
of space, which definitely leads to a kind of averaging of
the photon distribution $u(\vec R)$ details. So the details are not so sharp
as in the plane front case. Their characteristic width should be determined
by $0.5R$, and not by the distance of the star from the line of sight.
Calculations show that only the stars located inside or close to the
$1/\Gamma$ cone can produce sharp narrow peaks
(Fig.~\ref{OneStarProfiles}).

It is important that in the case of spherical jet geometry the peak is
asymmetric -- its tail is much longer that the front -- which hardly
corresponds to the observer profiles of gamma ray bursts.

In order to make the computed profiles  consistent with the observed ones,
the average distance between stars in the vicinity of the jet should be
\begin{equation}
\Delta R \sim \frac{c}{1-B}\Delta t
\end{equation}
where $\Delta t$ is the average time between the peaks in the burst. For
$\Gamma=1000$ $\Delta R \sim 0.02\Delta t$~pc/sec. It means that for realistic
gamma ray burst profiles the star density should exceed
\begin{equation}
\rho_\star \sim \Delta R^{-3} \sim 10^5 \quad\mbox{pc}^{-3}
\end{equation}
Such high density can occur only in globular clusters
and in the galactic nuclei.
The size of the region where the burst is formed should be of the order
$1$~pc which is similar to that of the globular cluster core.

After the above the gamma ray burst profiles
can be simulated from the integrating of the
contributions of all the stars in the vicinity of the jet. We assumed
that in the globular cluster core $10^6$ stars are located within $1$~pc$^3$.

\begin{figure*}
\epsfxsize=10cm
\centerline{\epsfbox{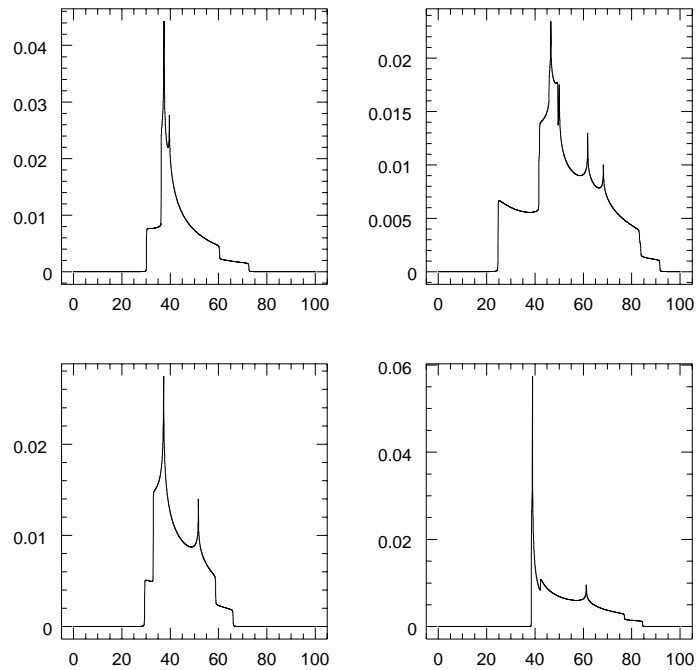}}
\caption{The example simulated gamma ray bursts profiles computed
 with a real distribution of stars in a globular cluster.
}
\label{SimulatedProfiles}
\end{figure*}

\section{The event rate}
In order to bind the gamma ray bursts to any events at cosmological
distances the rate density of such events should be
${\cal R}\sim 10^{-8}$~Mpc$^{-3}$~yr$^{-1}$. If we allow for a beamed radiation,
then ${\cal R}\sim 2\cdot 10^{-3} \beta^2_\circ$~Mpc$^{-3}$~yr$^{-1}$, where
$\beta_\circ$ is the beam width in degrees.

The rate of neutron star mergers in the dense stellar regions -- globular
clusters and galactic nuclei -- is
still not clear.

Numerical simulations of the evolution of close binaries
(Lipunov et al, 1996) show that
the neutron stars merger rate is strongly dependent on the stellar
 population age.
For the old stars of the globular clusters it should not exceed
$10^{-5}$~yr$^{-1}$ per $10^{11} M_\odot$; for the relatively young stars
in the galactic nuclei it should be of the order of
$10^{-4}$~yr$^{-1}$ per $10^{11} M_\odot$.
After taking into account the fact that the mass fraction of these
dense star configurations in the galaxies is not more than $10^{-2}$
we obtain the optimistic upper limit of about
$r = 10^{-6}$~yr$^{-1}$ per $10^{11} M_\odot$ galaxy.

For the flat Universe with $\Omega_{stars}=0.005$ and $H_0=75$~km/c/Mpc it
corresponds to the rate density ${\cal R}=10^{-8}$~yr~${-1}$. So, there is
no place for the beaming, and, thus, for the jets. At least, the
jets should be very wide (which supports the spherical geometry case described
in the previous section, in fact, if the jet front is plane, all emission should
be in very narrow $1/\Gamma$ beam).

If only the neutron star merger rate in globular clusters or galactic nuclei
is essentially higher than
$10^{-4}$~yr$^{-1}$ per $10^{11} M_\odot$
the above model can be used for explanation of the gamma ray bursts origin.

\section{Conclusions}
We have studied some of the conditions necessary for the explanation of
gamma ray bursts as the interaction of the baryonic
relativistic jet with the soft photon field in the dense star regions
(Shaviv, Dar 1996).
The main attractiveness of this model is that it is the only one that
predicts the observable gamma ray burst profiles.
We have found out two difficulties of these model, none of them being
fatal but each of them putting strong restrictions on it.
First of all, the main achievement of the model -- the reproduction
of the observational burst profiles -- has seemed to be strongly
sensitive to the jet front geometry. The shape of the peaks in the
spherical case becomes more wide and asymmetric, and the number of the peaks
becomes smaller.

The second is that it is hard to agree the theoretical event rate with
the observed one. The neutron star merger rate in globular clusters
and/or galactic nuclei should be high and the beaming of the jet
should be small.

The spherical geometry is more consistent with the event rate than the plane
one because it allows for beaming wider than $1/\Gamma$, but is in
worse agreement with the observed bursts profiles.
The detailed investigation of the jet expansion geometry would allow now
for the further treatment of this model of gamma ray bursts origin.

\begin{acknowledgements}
We acknowledge professor A. Dar for this exciting model of gamma ray bursts
and for interesting discussions of it. We are specially grateful to professor
V.M.~Lipunov and drs M.E.~Prokhorov and S.A.~Popov for useful advise and
discussions.
\end{acknowledgements}

\end{document}